\documentclass[%
reprint,
groupedaddress,
amsmath,amssymb,
aps,
prab,
floatfix,
]{revtex4-2}

\usepackage{graphicx}
\usepackage{dcolumn}
\usepackage{bm}
\usepackage{tikz}
\usepackage{pgfplots}
\pgfplotsset{compat=1.5}

\usetikzlibrary[arrows]
\usetikzlibrary{snakes}
\usetikzlibrary{decorations}
\usetikzlibrary {datavisualization.formats.functions}

\pgfdeclareradialshading{sphere1}{\pgfpoint{0.0cm}{0cm}}%
{rgb(0.0cm)=(0,0,1.0);
	rgb(0.5cm)=(0,1.0,0);
	rgb(1.0cm)=(1.0,0,0);
	rgb(1.02cm)=(0,0,0);
	rgb(1.3cm)=(0,0,0)}

\begin{document}
	
	\newcommand{\I}{\mathrm{i}}
	\newcommand{\E}{\mathrm{e}}
	\newcommand{\D}{\,\mathrm{d}}
	
	\title{Towards the use of X-ray Free-Electron Laser electron beams  to study Quantum Chromo-Dynamics} 
	\author{Eugene Bulyak}
	\email{bulyak@kipt.kharkov.ua, eugene.bulyak@desy.de}
	
	\affiliation{National Science Center `Kharkiv Institute of Physics and Technology', 1 Academichna str, Kharkiv, Ukraine\\ V.N.~Karazin National University, 4 Svodody sq., Kharkiv, Ukraine}
	\collaboration{LUXE Collaboration}
	
	\author{Svitozar Serkez}
	\email{svitozar.serkez@xfel.eu}
	\affiliation{European XFEL, Holzkoppel 4, 22869 Schenefeld, Germany}
	
	\author{Gianluca Aldo Geloni}
	\affiliation{European XFEL, Holzkoppel 4, 22869 Schenefeld, Germany}
	
	\date{\today}
	
\begin{abstract}
	X-ray free-electron lasers (XFELs) utilize high-density and high-energy electron bunches which are well-suited to produce Compton back-scattering radiation. Here we study back-scattered radiation pulses produced by the interaction of XFEL electron beams and  an optical laser. We discuss cost-effective setups to study such processes, taking advantage of the existing conventional as well as proposed XFEL infrastructure. We estimate parameters of possible experiments and compare them with other projects under construction.
\end{abstract}


\maketitle

\section{Introduction}
The purpose of X-Ray Free-Electron Lasers (XFEL) is to generate extremely brilliant, ultra-short and almost spatially coherent pulses of   X-rays at wavelengths down to the order of 1\,\AA, in order to exploit them for scientific experiments in a variety of disciplines spanning physics, chemistry, materials science and biology, \cite{nakatsutsumi2014}.

However, the electron  beams of XFELs possess unique characteristics in energy and phase-space density, which may also be applied for experiments in other areas of physics. In particular, if  these beams collide with intense pulses of quasi-monochromatic polarized  photons,  scattered high-energy polarized gamma-ray photons  are produced, which are extremely useful for paricle physics experiments and, in particular, Quantum Chromo-Dynamics (QCD).

Multi-GeV gamma-ray photons are useful tools in the investigation of the structures and properties of hadrons via their interaction  with a target nucleon or nucleus. In particular, they are suitable for hadron studies in the strange sector. The search for QCD exotics uses data from a wide range of experiments and production mechanisms. A few of such setups are currently under construction, see \cite{muramatsu22,adhikari21}. These setups do not cover the gamma energy range of 2.89--8.4 GeV.

Here we propose to employ the electron beams from the European XFEL accelerator to produce Compton back-scattered radiation pulses of polarized gamma-ray photons with the maximum energy of up to 4~GeV for QCD experiments.

After this introduction, the second section of this paper  briefly surveys properties of Compton back-scattering radiation with emphasis on  gamma-rays production. The third section presents a schematic setup of a Compton source based on the European XFEL electron beam to provide capabilities for QCD experiments  with photons in the multi--GeV energy range. Finally, the paper is concluded with a summary.       

\section{Compton back-scattering process at high electron energy}
Compton scattering, at variance with the classical Thomson process with  low energy  incident photons in the electron rest frame, exhibits modification of the spectra and angular distribution of the scattered photons due to recoil of the electrons.

\begin{figure}
	
	\begin{tikzpicture} [scale=1.5]
		
		\draw[->,red,thick] (0,0) -- (1,0) %
		node[above,red!50!gray] at (0.5,0.1) {electron}; %
		\filldraw[fill=red!30,opacity=0.5] (3,0) ellipse (0.5cm and 1cm);
		\filldraw[fill=red!30,opacity=0.5] (3,0) ellipse (0.25cm and 0.5cm);
		\filldraw[fill=red!30,opacity=0.5] (3,0) ellipse (0.1cm and 0.2cm);
		
		\draw[blue,snake=coil,segment aspect=0,->] (3,-2) -- (1,0);
		\node[below,blue] at (3,-2) {incident photon};
		
		\draw[black,thick] (1,0) -- (2.5,0);
		
		\draw[black,thick,opacity=0.4] (2.5,0) -- (3,0);
		
		\draw[black,thick,->] (3,0) -- (3.8,0);
		
		\draw[->] (1,0) -- (2.8,0.92); \draw[->] (1,0) -- (2.8,-0.92);
		
		\draw[opacity=0.7,->] (1,0) -- (2.92,0.49);
		
		\draw[opacity=0.7,->] (1,0) -- (2.92,-0.49);
		
		\draw[<->] (2,0) arc (0:28:1cm) node[above] at (1.7,0.5) {$\psi$};
		\node[above] at (3,1.1)  {scattered photon};
		
		\draw[<->] (2.2,0) arc (0:-40:1.2cm) node[below] at (1.7,-0.7)
		{$\phi$};
		
	\end{tikzpicture}
	
	\caption{Scheme of scattering the laser pulse.
		\label{fig:compscheme}}
\end{figure}
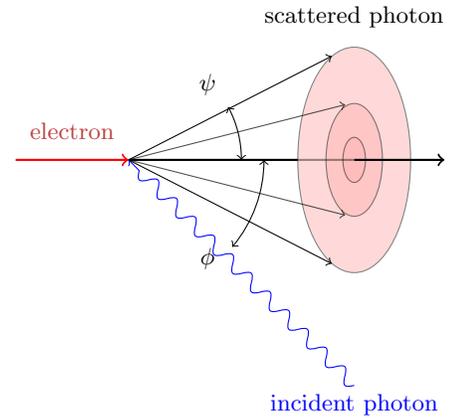

A scheme  of Compton back-scattering process  is presented in Fig.~\ref{fig:compscheme}.

Usually, Compton backscattered photons are referred  to as quasi-monochromatic. In fact, due to the strong correlation between the  scattering angle $\psi $ and the photon energy, proper collimation enables one to select a well-defined energy range. 

\subsection{Kinematics}
The relation between  the energy of a gamma-ray photon and the angle of emission determined from kinematics -- the energy-momentum conservation law -- reads (see, e.g., \cite{DANGELO20001,heinzl13}):
\begin{align}\label{eq:energy}
	\omega & \approx \frac{2\gamma^2_* (1+\cos \phi)\omega_\text{las}}{1+\gamma^2_*\psi^2} \; ; \\
	&\gamma_* := \frac{\gamma}{\sqrt{1+2(1+\cos \phi)\gamma\omega_\text{las}}} \;,
	\label{eq:gammast}
\end{align}
where $\omega $ is the energy of the scattered quanta, $\omega_\text{las}$ is the energy of laser photons,  $\gamma $ is the energy of electrons, $\phi$ and $\psi\ll 1 $ are the crossing angle and the scattering angle of photons, see Fig.~\ref{fig:compscheme}.

Here we make use the natural system of  units for particle and atomic physics, $\hbar = c = m_e = 1$. Therefore $\gamma := E_e/m_ec^2$ is the Lorentz-factor of electrons, $\omega := \epsilon/ m_e c^2$ equivalent Lorentz-factor of the scattered photons of energy $\epsilon$,  $\omega_\text{las} := \hbar \omega_\text{las}/ m_e c^2$ the equivalent Lorentz factor for the laser photons.

The energy of the recoiled electron becomes
\[
\gamma' = \gamma - \omega = \gamma\,\left( \frac{1+\gamma^2\psi^2}{1+2\gamma\omega_\text{las}(1+\cos\phi)}\right)\; ,
\]
and the energy of laser photon in the electron rest frame, see \cite{abramowicz21,fedotov23} is
\begin{equation} \label{eq:eta}
	\eta = (1+\cos \phi )\gamma \omega_\text{las}\approx 2 \gamma \omega_\text{las}\;  .
\end{equation}

\begin{figure} 
	\includegraphics[width=\columnwidth]{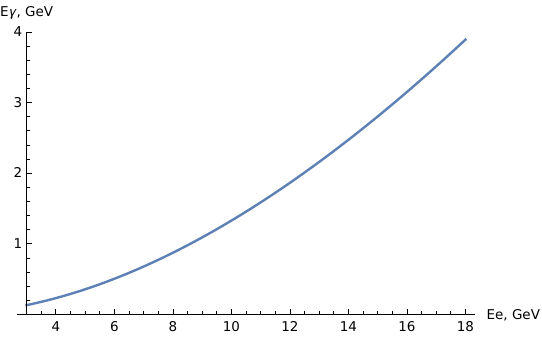}
	\caption{Maximum energy  of back-scattered gammas vs.  energy of electrons. \label{fig:Egvsund}}
\end{figure}

Under the assumption of small  angles, $\phi\ll 1, \psi\ll 1$, Eq.~\eqref{eq:energy} yields a simple expression for the maximum energy of the scattered quanta:
\begin{equation}
	\omega ^\mathrm{max} = \gamma \frac{4\gamma\omega_\text{las}}{1+4\gamma\omega_\text{las}}\;,
\end{equation}
which reveals that the recoil effect is governed by the term $4 \gamma \omega_\text{las} = 2\eta $.  For relatively weak electron recoil, $4 \gamma \omega_\text{las}\ll 1$, the maximum energy of the scattered  photons scales linearly with the laser photon energy and quadratically with the electron energy,
\[
\omega \approx 4 \gamma^2 \omega_\text{las}\, ,
\]
as illustrated in Fig.~\ref{fig:Egvsund} that is computed for the energy range  of the European XFEL accelerator  and laser photons with an energy of 1~eV. 

Accordingly, the \emph{minimal energy} of recoiled electron is
\[
\gamma'_\text{min} = \frac{\gamma}{1+4 \gamma\omega_\text{las}}\; .
\]

Another parameter  deduced from kinematics is the angle (measured from the incident electron trajectory) at which scattered photons possess the half-maximum energy, $\omega_{1/2} = \omega_\text{max}/2$:
\begin{equation} \label{eq:half}
 \psi_{1/2}= \frac{\sqrt{1+2\gamma \omega_\text{las}(1+\cos \phi)}}{\gamma} = \frac{1}{\gamma_*}\; .
\end{equation}

In the low-energy Thomson's limit, $\gamma_*\approx \gamma$, and the energy of gamma-rays drops by half at the emission angle of $1/\gamma $. In the electron rest frame it corresponds to the scenario when a photon scatters perpendicularly to the incident direction.

\subsection{Spectrum and cross-section}
A high-energy approximation for the Klein-Nishina  formula was derived by Arutyunian and Tumanian \cite{arutyunian63} (see also \cite{hajima2016}). One has the following dependence of the cross section on the frequency of the scattered photons:
\begin{widetext}
	\begin{align} \label{eq:spectr}
		\frac{\D \sigma_\text{Com} }{\D \omega} &=
		\frac{3 \sigma_\text{Th}}{16\gamma^2 \omega_\text{las}}\left[\frac{\omega^2}{4\gamma^2 \omega_\text{las}^2(\gamma-\omega)^2}-\frac{\omega}{\gamma \omega_\text{las}(\gamma-\omega)}+\frac{\gamma-\omega}{\gamma}+\frac{\gamma}{\gamma-\omega}\right]\mathrm{\Theta}\left(\frac{4\omega_\text{las}\gamma^2}{4\omega_\text{las}\gamma+1}-\omega\right)\; ,\\
		\intertext{while the integral is}
		\sigma_\text{Com} &=
		\frac{3  \sigma_\text{Th}}{32\gamma^3\omega_\text{las}^3}\left\{\frac{4\gamma \omega_\text{las}(2\gamma \omega_\text{las}(\gamma \omega_\text{las}+4)(2\gamma w+1)+1)}{(4\gamma \omega_\text{las}+1)^2}+\left[ 2\gamma \omega_\text{las}\gamma \omega_\text{las}-1)-1\right] \log(4\gamma \omega_\text{las}+1)\right\}\; , \label{eq:comcs}
	\end{align}
	where $ \sigma_\text{Th} = 8\pi r_0^2/3$ is the Thomson cross section ($r_0$ the classical electron radius), $\mathrm{\Theta}(\cdot ) $ is the Heaviside's Theta function (unit step).  
\end{widetext}

The Compton radiation spectrum given by Eq.~\eqref{eq:spectr} depends on the energy of the incident laser photon measured in the electron reference frame, $2\gamma\omega_\text{las}$. The total cross section, Eq.~\eqref{eq:comcs}, decreases with the energy of the incident laser photon. 

\begin{figure} 
	\includegraphics[width=\columnwidth]{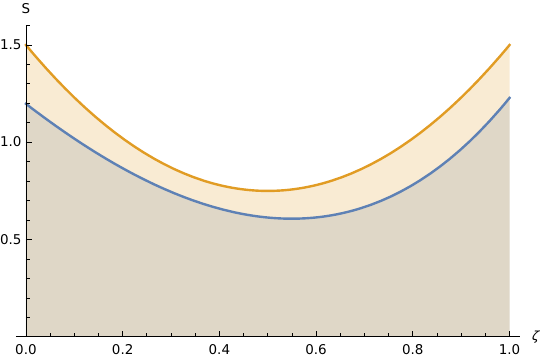}
	\caption{Rescaled spectra of Compton radiation $S$ defined in Eq.~\eqref{eq:spectrum} for an electron energy $E_e = 16.5\,\text{GeV}$ and an incident laser photon energy of 1~eV (blue curve), compared with the Thomson spectrum (orange curve). $\zeta = \omega / \omega_\text{max}$. The integral  Compton cross section is about 0.81 times the integral  Thomson cross section).  \label{fig:specs}}
\end{figure}

\subsection{Polarization of gammas} 
\begin{figure}
	\begin{tikzpicture}[scale=0.85] 
		\datavisualization [ scientific axes, x axis = {length=8cm,label={$\zeta = \omega / \omega ^\mathrm{max}$}, ticks=few},
		y axis={length=5 cm,include value=1.6,label={magnitude}},
		visualize as smooth line/.list={pos,tot,neg},
		style sheet=strong colors]
		data [set=tot,format=function] {
			var x : interval [0:1];
			func y = (\value x*\value x ))*3/2;
		}
		data [set=pos,format=function] {
			var x : interval [0:1];
			func y = (1-2* \value x*(1-\value x ))*3/2;
		}
		data [set=neg,format=function] {
			var x : interval [0:1];
			func y =(1-\value x)*(1-\value x )*3/2;
		};
	\end{tikzpicture}	
	\caption{Thomson photon spectrum (black) and its constituents: `positively polarized' (red) and `negatively polarized' (blue) gamma-ray photons. \label{fig:spectrum}}
\end{figure}
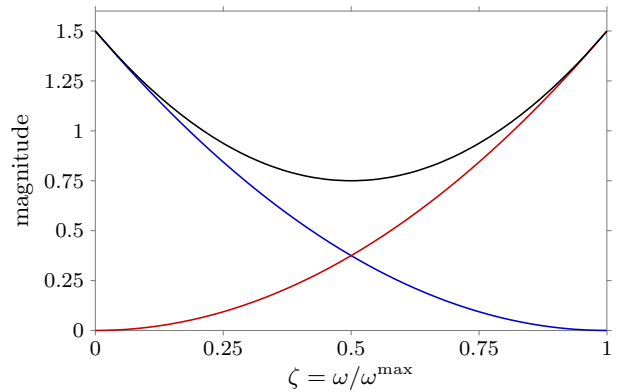

As is seen from Fig.~\ref{fig:specs}, for low energies of laser photons $\epsilon_\text{las}\lesssim 10\,\text{eV} $ 
the spectrum of Compton photons resembles the Thomson spectrum, which   has a simple form (see Fig.~\ref{fig:spectrum}  and \cite{bulyak05}):
\begin{equation}\label{eq:spectrum}
	S(\zeta) = \frac{3}{2}\left[ 1-2 \zeta (1-\zeta)\right]; \quad 0 \le \zeta\le 1\; ,
\end{equation}
where $\zeta := \omega / \omega^\text{max}$ is the energy of the emitted gamma-ray photon normalized to its maximum. The low-energy end of the spectrum, being of little significance for gamma sources,  scales as $\omega ^\text{min} \sim \omega_\text{las}/2\gamma^2$ and is approximated with zero.

The spectrum described by Eq.~\eqref{eq:spectrum} may be decomposed into a sum of two sub-spectra with `orthogonal' polarization components, see \cite{bulyak14a}. For a circular laser polarization, they are $S_\text{neg}$ -- the negative polarization constituent with the same polarization as the incident radiation, and $S_\text{pos}$ -- the positive constituent with the opposite polarization, see Fig.~\ref{fig:spectrum}:
\[
S_\text{pos}=3 \zeta^2/2\; ;\quad  S_\text{neg} = 3 (1-\zeta)^2/2\; .
\]
The degree of polarization $\rho$ is defined as
\[
\rho (\zeta )  := \frac{S_\text{pos}-S_\text{neg} }{S_\text{pos}+S_\text{neg}} =\frac{3\zeta}{2-\zeta+2\zeta^2}\; . \]

It should be emphasized that the  degree of polarization has probabilistic sense: if, for instance, $\rho = 0.7$, one obtains 85\% of gamma-ray photons `positively polarized' and 15\% `negatively polarized'.  

\begin{figure}	
\includegraphics[width=\columnwidth]{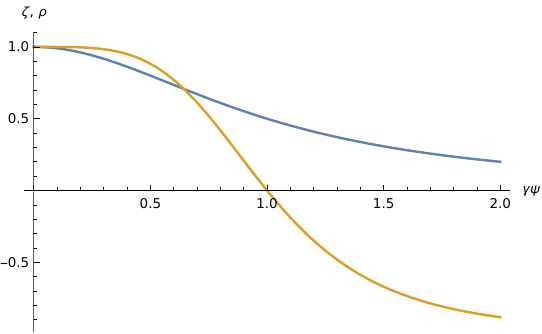}
\caption{Relative energy of gamma ray photons (blue curve) and  degree of polarization (orange) vs.  emission angle. \label{fig:enpol}}
\end{figure}

The dependence of the degree of polarization $\rho$ and of the normalized gamma-ray photon energy $\zeta$ as a function of the emission angle (multiplied by $\gamma_*$) is presented in Fig.~\ref{fig:enpol}.

As it can be seen from the figure, at the angle $\gamma_* \psi =1$, the photon energy decreases by half, and the  degree of polarization is zero, i.e. positively- and negatively-polarized photons are equal in numbers. Upon further increase of the observation angle, the photon energy continues to drop while the number of negatively polarized photons prevails over the positive ones.

 Compton radiation is quasi-monochromatic: if the laser beam is scattered by a zero-emittance electron beam and observed through an angular filter (a pinhole or a mask selecting a narrow ``ring'' around the radiation axis that matches the electron trajectory)  , every  photon scattered by an electron with a given energy at a definite angle to its trajectory has a fixed energy and  degree of polarization. 	
 
In contrast to a single electron or a monoenergetic filament electron beam, realistic electron beams with finite energy spread and emittance, scatter-off radiation with larger spectral bandwidth; at XFELs the contribution of emittance is dominant~\cite{bulyak14a}.
In the case of multi-GeV electrons the energy half-maximum angle (zero degree of polarization)  $\psi_{1/2}$ is rather small. For example,  for $E_e = 16.5\,\text{GeV}$ and $\epsilon = 1\,\text{eV}$ the photon energy is diminished by half at  $\psi_{1/2} \approx 35\,\mu\text{rad}$.  

 The  correlation of the photon  degree of polarization with its energy is independent of electron's trajectory direction. A Coincidence scheme for defining the gamma-ray photon energy and polarization through measurement of the electron energy that emitted it is therefore justified.

\subsection{Yield of photons}
The yield of gamma-ray photons  per single crossing of the incident photon pulse with the electron bunch, both assumed having a 3D Gaussian shape \cite{bulyak05},  reads
\begin{equation}\label{eq:yield}
	Y = \frac{N_\text{pp}N_e\, \sigma_\text{C} }{2\pi
		\sqrt{\left({\sigma'}_z^2+\sigma_z^2\right)\left[\sigma_x^2+{\sigma'}_x^2+\left( \sigma_y^2
			+{\sigma'}_y^2\right) \tan^2\frac{\phi}{2} \right]} }\; ,
\end{equation}
where $N_\text{pp}, N_e$ are populations of the incident radiation pulse and of the bunch, respectively, $\sigma_{x,y,z}, \sigma'_{x,y,z}$ are their horizontal, longitudinal, and vertical rms dimensions and $\sigma_\text{C} $ is the total Compton cross-section  in Eq.~\eqref{eq:comcs}. 

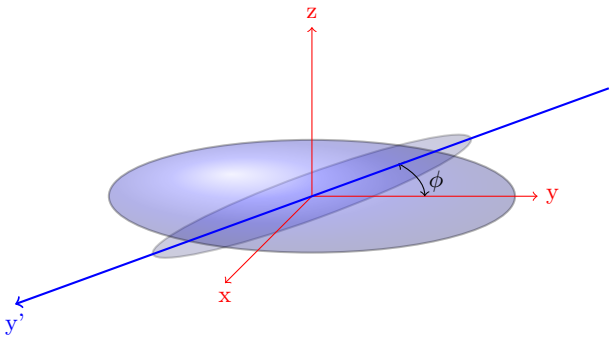
\begin{figure}
	\begin{tikzpicture}[scale=1.5]
		\shadedraw [rotate=20,shading=ball,opacity=0.2,thick] (0,0) ellipse
		(1.5cm and 0.2cm);
		
		\shadedraw [shading=ball,opacity=0.3,thick] (0,0) ellipse (1.8cm and 0.5cm);
		
		\draw[->,red] (0,0,0) -- (2,0,0) node[right] {y}; %
		\draw[->,red] (0,0,0) -- (0,1.5,0) node[above] {z}; %
		\draw[->,red] (0,0,0) -- (0,0,2) node[below] {x};
		
		\draw[<->] (1,0) arc (0:40:1cm and 0.45cm); %
		\node[above] at (1.1cm,-1pt) {$\phi$};
		
		\draw[->,rotate=20,blue,thick] (2.8,0) -- (-2.8,0) node[below] {y'};
		
	\end{tikzpicture}
	\caption{Scheme of bunch--pulse crossing. \label{fig:scheme}}
\end{figure}	

\section{Compton sources at the European XFEL}
Owing to the high density and high energy electron beams needed to drive X-ray FELs, we propose to adapt them as Compton sources  of gamma-ray photons in the GeV energy range,  e.g., for experiments in Quantum Chromo-Dynamics. 

\begin{figure}
	\begin{tikzpicture} [scale=0.5]
		\filldraw[fill=gray!40]  (0,0) rectangle (5,4) (2.5,4.3) node[black]{magnet spectrometer } ;
		
		\draw[red,very thick] (-6,0.5) node[above]{electrons} -- (0,0.5) arc[radius=10,start angle =270,delta angle=30] -- (8.5,4) node[above,rotate=33]{to dump};
		
		\foreach \x / \angle / \y in { 9/33.7/3.0, 8/38.7/3.5, 7/45.6/4.25, 6/56.44/5.43}
		\draw[red!40!gray] (0,0.5)  arc[radius=\x,start angle =270,delta angle=\angle] -- (6.5,\y);
		\draw[white, very thick] (-4,0.48) -- (8,0.43) -- (8,0.53) -- cycle;
		\filldraw[blue!80] (-4,0.48) -- (8,0.43) -- (8,0.53) -- cycle 
		(6.5,0.6) node[blue,above]{photons};
		\filldraw[yellow!60!black] (8,0) node[below,yellow!40!black]{target} rectangle (8.5,1) node[black,rotate=90,right]{coinc.};
		
		\filldraw[blue!40]  (-4,0) rectangle (-3.5,1.0) (-3.5,2) node[black,rotate=90,above]{radiator}; 
		\filldraw[blue!50!red]  (-3.75,0.48) ellipse [x radius=4pt, y radius=2pt]; 
		
		\filldraw[green!60!black] (6.5,3) rectangle (6.6,5.5) node[above]{tagger};
		\draw[thick,out=90,in=330,<<->>] (8.25,1)   to  (6.6,5);
	\end{tikzpicture}
	\caption{Scheme of experimental setup. \label{fig:scheme1}}
\end{figure}
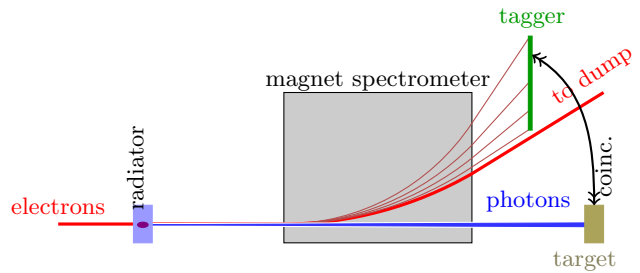

Figure~\ref{fig:scheme1} presents a scheme of such experimental setup.  It is important  that the electron spectrum upon interaction with the radiation pulse consists of two distinct parts: the recoiled and the undisturbed fractions. The recoiled fraction is small, less than $10^{-4}$ of the total. 

\begin{figure}
	\centering
	\begin{tikzpicture}[scale=1.75]
		\shade[left color=gray!40,right color=blue] (0,0) -- (0,1.5) parabola bend (1,0.75) (2,1.5) |- (0,0);
		\draw[->] (-0.1,0) -- (2.2,0) node[below] {$\epsilon_\gamma$};
		\shade[left color=gray!40,right color=red!50] (2.5,0) -- (2.5,1.5) parabola bend (3.5,0.75) (4.5,1.5) |- (2.5,0);
		\draw[red] (4.6,0) node[right,rotate=90]{non-scattered electrons} ;
		\draw[->] (2.4,0) -- (4.8,0) node[below] {$E_\text{el}$};
		\draw[thick,out=100,in=80,<->,green!40!black] (1.9,1.4)   to  (2.6,1.4); 
		
		\draw[thick,out=90,in=90,<->,] (1,0.8)   to  (3.5,0.8); 
		\draw (2.25,1.6) node[above]{coincidence} ;
		
	\end{tikzpicture}
	\caption{Relation of the electron's (red) to gamma's (blue) spectra. \label{fig:elphot}}
\end{figure}
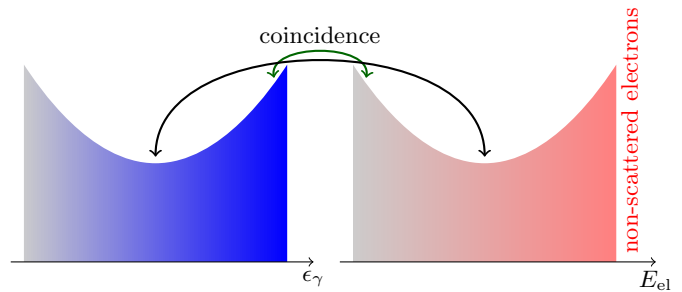

`Photon tagging' assigns  the energy loss of \emph{individual electron}  to the energy of the emitted photon in coincidence, as depicted in  Fig.~\ref{fig:elphot}. 

\subsection{Radiation source: infrared laser}
Below we discuss whether a ten $\mu$A - order  average current, deliverable by the accelerator of the European XFEL, is able to produce sufficient flux of scattered GeV photons.

The number of back-scattered photons per crossing  is proportional to the populations  and a geometrical factor. The latter  is inversely proportional to the dimensions of bunch and photon pulse. The bunch population is by three orders of magnitude smaller than that in LEPS (storage ring SPring-8, \cite{muramatsu22}), but it is compensated  by a much larger population of the laser photons stored in resonator. In addition, the small transverse size of low-emittance electron bunches at the European XFEL enhances the yield as well.


The key element of the laser system proposed for the European XFEL is a resonator. The resonator is capable of storing up to 10--20 thousand pulses of 100~Watt infrared laser. The stored energy may reach levels as high as 30~mJ per pulse, \cite{dupraz20a}.

The main parameters of the proposed gamma source together with those under construction, LEPS2 \cite{muramatsu22} and GlueX \cite{adhikari21}, are listed in Table~\ref{tab:eubased}.
The range of gamma-photon  energy for the European XFEL-based source is taken such that the polarization at the bottom edge  equals to 1/2: 
\[
\rho (\zeta_{1/2} )=\frac{1}{2}\Rightarrow \zeta_{1/2} =\frac{3-\sqrt{3}}{2}\approx 0.63\;  .
\] 
The  polarization is linear for GlueX, while for other projects it is determined by the polarization of laser. 
Corresponding parameters in each column are marked with the same color.

\begin{table}
	\caption{Main parameters of gamma sources.  \label{tab:eubased}}
	\begin{ruledtabular}
		\begin{tabular}{lrrr}
			\hfill experiment & LEPS2 & GlueX & XFEL+laser\\
			\hfill parameter  & (SPring8)&(CEBAF)& (European XFEL) \\
			\hline
			$E_e$, GeV & 8 & 12 & \textcolor{red}{4.0}  \\
			&   &  & 8.0  \\
			&   &  & 11.5 \\
			&   &  & 14.0  \\
			&   &   & 16.5 \\
			&   &   &\textcolor{blue}{17.5}  \\
			\hline
			$I_\text{beam}$, A &0.1 & $1.5\times 10^{-7}$ & $ 2.7\times 10^{-5}$  \\
			\hline
			radiator & lasers & diamond & las+res  \\
			& & $\Delta =50\,\mu\text{m}$ & \\
			$\lambda_\text{las}$, nm &\textcolor{red}{355}& $\sim 2\times  0.15$& 1064 \\
			&\textcolor{blue}{266}& &\\
			$\epsilon_\gamma$, GeV &\textcolor{red}{1.3--2.39} &8.4--9&\textcolor{red}{0.15--0.23} \\
			&\textcolor{blue}{1.6--2.89}&&0.55--0.87 \\
			&   & &1.09--1.72\\
			&  &  & 1.57--2.47\\
			& & & 2.11--3.33\\
			&  & &\textcolor{blue}{2.62--4.14} \\
			$\rho$, \% & $\lesssim 98$& (35--40)& 50--100 \\
			$W_\text{laser}$, W &8& & $\lesssim 10^6$ \\
			&0.8 & & \\
			flux, phot/s &\textcolor{red}{$10^6$}&$10^7 $&$ \lesssim  10^9$~\footnote{assuming $10^3$ events per second} \\
			&\textcolor{blue}{$10^5$}&& \\
		\end{tabular}
	\end{ruledtabular}
\end{table}

%

\subsection{Capability of XFEL accelerator to produce GeV photons}
Comparing the European XFEL accelerator to the storage ring SPring-8, \cite{muramatsu21}, where the GeV-photons Compton source LEPS2 has been built, one can see a  4 orders of magnitude larger  current of the electrons circulating in the ring,  $\sim 300\,\text{mA} $, as compared with   $\sim 10\,\mu\text{A}$ in the linac. On the other hand, due to a factor two  higher maximum electron energy of the linac (17.5\,GeV), the European XFEL could produce four times higher  photon energy with the same laser wavelength. Moreover, the quality of the linac beam -- in terms of transverse emittances and the bunch length -- allows for substantial reduction of the geometric factor [the denominator in \eqref{eq:yield}].  

Finally, for the realistic case of a moderate but still higher energy than attained in \cite{muramatsu21}, the laser resonator of the kind developed for Compton x-ray sources -- see, e.g., \cite{bruni11,dupraz20,gunther20} -- and the  sources of polarized positrons for ILC and CLIC projects of lepton colliders \cite{gladkikh13,bulyak15d,bulyak22b} may be employed.
The optical resonator is capable of increasing the power of a YAG laser (1\,$\mu$m wavelength) by $(2\dots 4)\times 10^3$ times, see e.g.  papers on Compton X-ray compact sources \cite{brummer22,hornberger21}. 

Making use of the parameters of the European XFEL accelerator \cite{xfelacc}, we estimated the performance of a Compton source of polarized GeV photons. We took the parameters of a laser system that was considered for the projects of lepton colliders ILC and CLIC. Currently, a similar system is under construction at the IJC laboratory (Orsay, France) for the ThomX Compton source. 

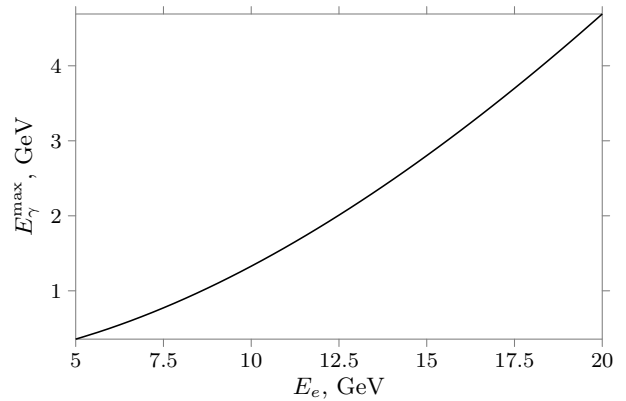
\begin{figure} 
	\begin{tikzpicture} [scale=1.4]
		\datavisualization [scientific axes, x axis = {label={$E_e$, GeV}},
		y axis = {label={$E^\mathrm{max}_\gamma$, GeV}},
		visualize as smooth line]
		data [format=function] {
			var x : interval [5:20];
			func y = \value x*\value x /(\value x + 65.28);
		};
	\end{tikzpicture}
	\caption{The  gamma-photon maximal energy as a function of the electron energy for $E_\text{laser} = 1\,\text{eV}$. \label{fig:maxenergy}}
\end{figure}

The dependency of the maximum energy of photons on the electron energy is presented in Fig.\ref{fig:maxenergy}.

A 16.5\,GeV electron beam of the European XFEL, assuming a laser resonator storing 1\,mJ laser pulse at 1.164\,eV photon energy (providing 8 degrees crossing angle) produces a Compton spectrum with the maximum energy $\sim 4$\,GeV with an efficiency of $ 10^{-4}$ photons per electron-crossing. The data from estimations and simulations are listed in Table~\ref{tab:numbers}. 

\begin{figure}
\begin{tikzpicture}[scale=0.85] 
	\datavisualization [scientific axes, x axis = {length=8 cm,label={$\zeta = \omega / \omega ^\mathrm{max}$ }}, y axis = {length=5 cm,include value=100,label={photons/bin}}, visualize as  line/.list={tot,pos,neg},
	style sheet=strong colors]	
	
	data [set=tot] {
		x, y
		0.400,    59
		0.444,     59
		0.444,     33
		0.488,     33
		0.488,     36
		0.531,      36
		0.531,      54
		0.575,      54
		0.575,      47
		0.619,      47
		0.619,      54
		0.663,     54
		0.663,     60
		0.706,     60
		0.706,     62
		0.750,     62
		0.750,     63
		0.794,     63
		0.794,     71
		0.837,     71
		0.837,     77
		0.881,     77
		0.881,     63
		0.925,     63
		0.925,     90
		0.969,     90
		0.969,     66
		1.013,      66
		1.013,     0
		1.056,     0
	}
	data [set=pos] {
		x, y
		0.400,    21
		0.444,     21
		0.444,     22
		0.488,     22
		0.488,     20
		0.531,      20
		0.531,      30
		0.575,     30
		0.575,     35
		0.619,     35
		0.619,     43
		0.663,    43
		0.663,    56
		0.706,     56
		0.706,     51
		0.750,     51
		0.750,     59
		0.794,     59
		0.794,     68
		0.837,     68
		0.837,     77
		0.881,     77
		0.881,     63
		0.925,     63
		0.925,     90
		0.969,     90
		0.969,     66
		1.013,      66
		1.013,      0
		1.056,     0
	}
	data [set=neg] {
		x, y
		0.400,   38
		0.444,   38
		0.444,    11
		0.488,    11
		0.488,    16
		0.531,     16
		0.531,     24
		0.575,     24
		0.575,     12
		0.619,     12
		0.619,     11
		0.663,    11
		0.663,     4
		0.706,     4
		0.706,    11
		0.750,    11
		0.750,     4
		0.794,     4
		0.794,     3
		0.837,     3
		0.837,     0
		0.881,     0
		0.881,     0
		0.925,    0
		0.925,    0
		0.969,    0
		0.969,    0
		1.013,     0
		1.013,    0
		1.056,    0
	};
\end{tikzpicture}

\caption{Simulated photon spectrum (black) and its constituents: `positively polarized gammas' (red) and `negatively polarized' (blue). 16.5~GeV electrons crossed by 1.164~eV laser at 8 deg. \label{fig:colspectrum}}
\end{figure}
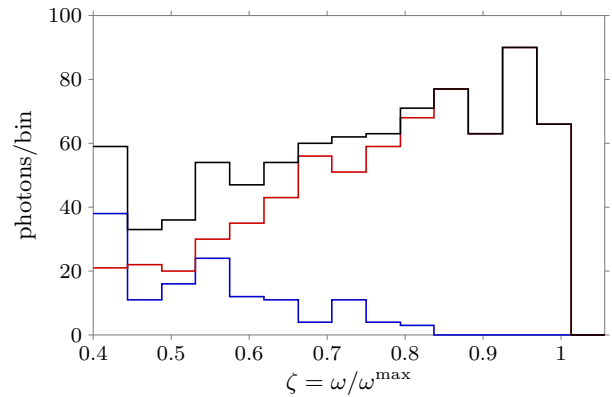

Simulations were done with a Monte--Carlo code initially written for study of beam dynamics in Compton storage rings.
Spectra of the photon flux are presented in Fig.~\ref{fig:colspectrum}.

\begin{table}\label{tab:numbers}
	\caption{Simulated parameters of Compton source}
	\begin{ruledtabular}
	\begin{tabular}{lrl}
		parameter & value &units \\ \hline 
		electron energy & 16.5 & GeV \\ 
		bunch charge & 1 & nC \\
		bunch length & 24 & $\mu$m \\
		bunch tr. $\sigma_x/\sigma_z$ at IP & 20/20 &$\mu$m \\
		\hline
		energy of photon& 1.164 & eV \\
		pulse energy & 1 & mJ \\ 
		pulse trans. $\sigma'_x/\sigma'_z$ at IP & 20/20 &$\mu$m \\
		pulse long. $\sigma'_y$ at IP & 50 & $\mu$m \\
		crossing angle & 8  & deg \\
		\hline
		max. gammas energy &3.88 &GeV \\
		half max angle $\psi_{1/2}$ &35 & $\mu$rad \\
		yield per crossing & $3.5\times 10^4$ & gammas \\
	\end{tabular} 
	\end{ruledtabular}	
\end{table}

Our simulations show that the spectrum is rather uniform, and the fraction of negatively polarized gammas negligible for $\zeta \gtrsim 0.8$. 

\section{Summary and Discussion}
Installation of a laser resonator and an electron tagger system at the European XFEL makes it possible to carry out QCD experiments. The energy range of Compton backscattered gamma-ray photons, their flux and polarization  are in range with those at experimental setups under construction  and fill the `energy gap' between the most modern experimental setups .

Realization of the proposal would be a ``stepping stone'' to LUXE project \cite{abramowicz21} dedicated to the study of interaction between relativistic electrons with powerful laser pulses.

\begin{acknowledgments}
	Authors would like to thank 
	Serguei Molodtsov for his interest and support. Computations were done with help of \textsc{Wolfram Mathematica Cloud} package \cite{Mathematica}, provided to one of coauthors (EB).
\end{acknowledgments}
\providecommand{\noopsort}[1]{}\providecommand{\singleletter}[1]{#1}%
\end{document}